\documentclass[iop,apj,12pt,preprint]{emulateapj}

\newcommand{\target}{GRS1915+105}
\newcommand{\instrument}{X-shooter~}

\slugcomment{Accepted for publication in the Astrophysical Journal}

\shorttitle{The black hole in \target~}
\shortauthors{Steeghs et al.}

\begin{document}


\title{The not-so-massive black hole in the microquasar \target}


\author{D.Steeghs\altaffilmark{1,2}}
%\affil{}
%\and
\author{J.E.McClintock\altaffilmark{2}}
%\and
\author{S.G.Parsons\altaffilmark{1,3}}
%\and
\author{M.J.Reid\altaffilmark{2}}
\author{S.Littlefair\altaffilmark{4}}
%\and
\author{V.S.Dhillon\altaffilmark{4}}
%\affil{}


\altaffiltext{1}{Department of Physics, Astronomy and Astrophysics group, University of Warwick, CV4 7AL, Coventry, UK}
\altaffiltext{2}{Harvard-Smithsonian Center for Astrophysics, 60 Garden Street, Cambridge, MA 02138, USA}
\altaffiltext{3}{Departamento de Física y Astronomía, Facultad de Ciencias, Universidad de Valparaiso, Chile}
\altaffiltext{4}{Department of Physics and Astronomy, University of Sheffield, S3 7RH, Sheffield, UK}


\begin{abstract}

We present a new dynamical study of the black hole X-ray transient \target~making use of near-infrared spectroscopy obtained with \instrument at the VLT. We detect a large number of donor star absorption features across a wide range of wavelengths spanning the H and K bands. Our 24 epochs covering a baseline of over 1 year permit us to determine a new binary ephemeris including a refined orbital period of $P=33.85 \pm 0.16$d. The donor star radial velocity curves deliver a significantly improved determination of the donor semi-amplitude which is both accurate ($K_2=126\pm1$ km/s) and robust against choice of donor star template and spectral features used. We furthermore constrain the donor star's rotational broadening to $v\sin{i}=21\pm4$ km/s, delivering a binary mass ratio of $q=0.042 \pm 0.024$. If we combine these new constraints with distance and inclination estimates derived from modeling the radio emission, a black hole mass of $M_{BH}=10.1 \pm 0.6 M_{\odot}$ is inferred, paired with an evolved mass 
donor of $M_2=0.47 \pm 0.27 M_{\odot}$. Our analysis suggests a more typical black hole mass for \target~rather than the unusually 
high values derived in the pioneering dynamical study by Greiner et al. (2001). Our data demonstrate that high-resolution infrared spectroscopy of obscured accreting binaries can deliver dynamical mass determinations with a precision on par with optical studies.

\end{abstract}





\section{Introduction}

Mass is the most fundamental property of a black hole.  It defines
both a length scale and a time scale and, in conjunction with the
black hole's spin, provides a complete description of any astronomical
black hole.  Knowledge of mass is essential in assessing how a black
hole interacts with its environment.  For example, it defines the
Eddington luminosity, the maximum isotropic radiative luminosity that
a black hole can achieve.  In this paper, we present accurate
dynamical data for the X-ray transient \target, including a significantly refined measurement of the mass of its
black hole primary.

Black hole X-ray transients are compact binaries where a stellar mass black hole accretes from a typically low-mass companion (Remillard \& McClintock 2006). They show complex and varied outburst behaviour, though they spend most of their life in a relatively low accretion rate state. In many respects, the most extraordinary such system is \target.  It has remained active since its eruption and discovery in 1992, while other black hole transients
return to quiescence after only a year or so following an outburst.
Not only has \target~remained active, it is frequently observed in super-Eddington states when it is the most luminous accreting black hole in the Galaxy ($L_{bol}\sim10^{39}$ erg/s, e.g. Done et al.\ 2004).  The
secondary is a K-type giant (Greiner et al. 2001a), and the binary system is
far larger than that of any other black hole transient. Its 34-day
orbital period is five times that of second-place V404 Cyg, and the
volume of the Roche lobe of its black hole is correspondingly fifty
times larger.  Concerning X-ray variability, \target~ is spectacularly unique. 
More than a dozen distinct X-ray states have been identified (Belloni
et al. 2000; Klein-Wolt 2002), and the source displays a constellation
of four high-frequency QPOs, rather than the customary one or two
(Remillard \& McClintock 2006).  As the prototype of microquasars, the
signature feature of \target~is its pc-scale bipolar radio jet
(Mirabel \& Rodriguez 1994; Fender et al.\ 1999). The jet is likely
powered by the spin of the black hole (Narayan \& McClintock 2012);
the spin is extreme ($a/M > 0.98$) and far higher than that measured for
other black hole transients (McClintock et al.\ 2006; Blum et al.\
2009).  

For the purposes of this paper, the most interesting contrast
between \target~and the other black hole transients is the greater
mass of its black hole primary.
The black hole mass of \target~was first measured by Greiner et al. (2001b, hereafter G01)
to be $M=14.0\pm4M_{\odot}$.  It is accurate to describe this
dynamical study by G01 as pioneering for two reasons. First, the observations were necessarily made in the near-infrared
because the system is inaccessible at optical wavelengths, with
$A_{\rm V}\approx20$~mag (Chapuis \& Corbel 2005; Rahoui et al.\
2010).  Second, this was the first successful dynamical study of a
black hole transient in an active state in the presence of intense
X-ray heating.  As has often been remarked, the nominal mass measured
by Greiner et al.\ is much greater than the masses of other black hole
transients, which fall in a relatively narrow range ($7.8 \pm 1.2 M_{\odot}$; Ozel et al. 2010, see also Farr et al. 2011). However, given the relatively large uncertainty on the black hole mass, this difference is not at high statistical significance.

For their dynamical study, G01 used the ISAAC NIR spectrograph on the VLT-Antu telescope. Hurley et al. (2013) recently revisited the ISAAC data together with  additional unpublished data.  Here we present higher spectral resolution data aimed at significantly improving the precision with which we can determine the black hole mass. Our data were obtained with the \instrument spectrograph that was recently commissioned on the VLT (Vernet et al. 2011). Thanks to its multi-arm echellette design, \instrument offers a much wider spectral bandpass at higher dispersion while S/N is excellent thanks to its high throughput. In Section \ref{obs}, we discuss our observational campaign and detail our data reduction and analysis in Section \ref{redu}. Our key results are presented in Section \ref{anal}, from which we derive new constraints on the mass of the black hole (Section \ref{mass}). We conclude our study in Section \ref{conclude}.

\section{Observations}
\label{obs}

\begin{table}
%\scriptsize
\begin{center}
\caption{Log of Observations\label{obslog}}
\begin{tabular}{llc}
\tableline\tableline
Date (UTC)   &   Object       & Exptime (s)  \\     
\tableline
13 May 2010 & HD179130 & 3\\
13 May 2010 & HD174336 & 3\\
6 June 2010 & HD180732 & 3\\
6 June 2010 & HD179691 & 6\\
8 June 2010 & HD176354 & 2\\
9 June 2010 & \target~ & 2400\\
9 July 2010 & \target~ & 2400\\
11 July 2010 & \target~ & 2400\\
13 July 2010 & \target~ & 2400\\
27 July 2010 & \target~ & 2400\\
2 Aug 2010 & \target~ & 2400\\
10 Aug 2010 & \target~ & 2400\\
19 Aug 2010 & \target~ & 2400\\
27 Aug 2010 & \target~ & 2400\\
19 Sep 2010 & \target~ & 2400\\
24 Sep 2010 & \target~ & 2400\\
7 Oct 2010 & \target~ & 2400\\
21 Apr 2011 & \target~ & 2400\\
1 May 2011 & \target~ & 2400\\
3 May 2011 & \target~ & 2400\\
7 May 2011 & \target~ & 2400\\
9 May 2011 & \target~ & 2400\\
13 May 2011 & \target~ & 2400\\
18 May 2011 & \target~ & 2400\\
13 June 2011 & \target~ & 2400\\
15 June 2011 & \target~ & 2400\\
24 June 2011 & \target~ & 2400\\
17 July 2011 & \target~ & 2400\\
4 Aug 2011 & \target~ & 2400\\
\tableline
\end{tabular}
%% Any table notes must follow the \end{tabular} command.
%\tablenotetext{a}{The two columns reflect our two choices for the linear limb darkening coefficient}
\end{center}
\end{table}

We employed the multi-arm \instrument spectrograph attached to the VLT-UT2 telescope at ESO Paranal Observatory to conduct our spectroscopic campaign of \target. All observations were executed in service mode. A key strength of the \instrument instrument is its ability to simultaneously observe both the full optical band as well as the near-infrared across its three arms at an intermediate spectral resolution and with high throughput (Vernet et al. 2011). Due to the high extinction towards \target, we did not anticipate detecting the source with the VLT shortwards of the J-band. Despite this, \instrument is still the ideal instrument for NIR spectroscopy of \target~ given its optimal compromise between high efficiency, sufficient resolution and full coverage of the NIR bands up to 2.48$\mu m$. 

All our observations employed a 0.6" slit to achieve a spectral resolution of R=8,000 in the NIR arm, representing a significant improvement over the R=3,000 spectra obtained with ISAAC (Greiner et al. 2001a; G01; Hurley et al. 2013). Seeing conditions were constrained to better than 0.8" in order to control slit-losses. At each epoch, three 15 minute exposures were obtained while nodding along the slit in between these exposures to mitigate sky background subtraction errors. Given that the \instrument acquisition camera operates in the optical band, where \target~ is extremely faint ($(I-K)\sim10$), the target was acquired using $z$-band (900nm) acquisition images, applying a fixed offset relative to a nearby field star to align the target on slit.

In total, 24 sets of exposures were obtained between June 9th 2010 and August 4th 2011 in service mode by the UT2 team. In addition, exposures of telluric standards were obtained close in time with each target visit and 5 bright K-M type giants were observed with the same setup to serve as spectral templates for the data analysis. See Table \ref{obslog} for a log of observations.

\section{Reduction}
\label{redu}

We reduced these data using version 1.3.7 of the X-shooter pipeline. The standard recipes were used to optimally extract and
wavelength calibrate both the target observations as well as our telluric and spectral standard stars. The 3 nodded sub-exposures for each epoch were combined before extraction to obtain a single good S/N spectrum while improving sky subtraction. The resultant spectra cover 9940-24790\AA~at 0.6\AA/pixel though \target~is only significantly detected longwards of 11500\AA.  The wavelength calibration step makes use of a daytime arc exposure. From these we also confirmed that our achieved spectral resolution was indeed R=8,000, delivering a resolution element of width 3.5 detector pixels.

Given that our data were obtained over a fairly long time-interval, we removed any residual wavelength shifts between epochs by forcing telluric features to align. This was achieved by cross-correlation of the same spectral segment in the K-band that is dominated by well-resolved tellurics. Spectra were first re-binned to a constant velocity-scale of 11 km/s/pixel in order to ensure that by shifting spectra for cross-correlation purposes, pixel shifts map to global velocity shifts. We found significant shifts  between spectra ranging from -10 to +13 km/s with an RMS of 5 km/s. While these shifts were only of the order of one pixel, measurement uncertainties were much less than 1 km/s thanks to the high S/N in the K-band tellurics. We also tried different telluric regions and found that the resultant shifts were always consistent within 1-2 km/s. We thus found that this step improved the stability of our wavelength scale across our campaign to better than 2 km/s. Once aligned, we then used the relevant 
telluric star observations closest to each target observation to perform a telluric correction by scaling with a continuum normalised version of the telluric. The telluric-corrected spectra were then shifted into a heliocentric velocity frame, making appropriate time and velocity corrections. All velocities and times quoted in this paper are thus in a heliocentric frame, using the UTC time system.
% added to warn against Balmer line profiles
We note that the available telluric stars were of early spectral type and different epochs were corrected with different telluric stars. Some artefacts are thus introduced in the target spectra due to absorption features in the tellurics themselves, mainly restricted to hydrogen absorption. This does not affect our analysis as we are only considering the absorption features of the late type donor in \target, but this does limit the ability to use our data to analyse any hydrogen emission profiles in \target.

\begin{figure}
\centerline{\includegraphics[width=9cm,clip='yes']{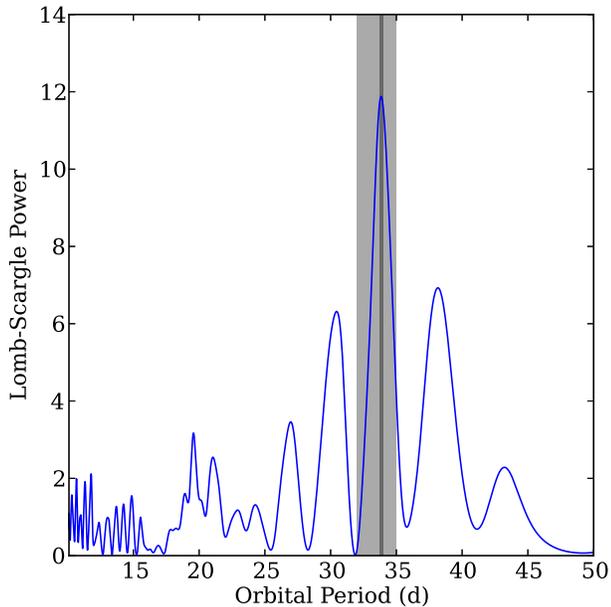}}
\caption{Lomb-Scargle periodogram obtained from our radial velocities spanning over 1 year. A prominent peak is observed at the orbital period of the binary. The dark shaded area denotes our binary period determination of $P=33.85\pm0.16$d, while the lighter gray area denotes the period constraint from Greiner et al. (2001).\label{scargleplot}}
\end{figure}

\section{Analysis}
\label{anal}

In order to determine radial velocities (RVs) of the donor star features in \target~at each epoch, we normalised our target and spectral template observations and again used the cross-correlation technique to determine the radial velocity shifts between target and template. As mentioned previously, the data were binned to a constant velocity sampling per pixel to facilitate this. Suitable masks were defined to exclude regions where \target~has emission features and to mask any residual telluric standard features. In addition, a significant number of hot pixels were present in the extracted spectra that had to be flagged and masked. This was an iterative procedure largely based on visual inspection as the severity of these hot pixels varied greatly from epoch to epoch. Some of these are introduced by anomalous pixels in the telluric template, others in the target spectrum. Masking is preferred over attempting to interpolate over such pixels/features as incorrect interpolation could lead to spurious velocities.

Several spectral regions were investigated to test the robustness of our velocities. 
The choice of these regions were largely motivated by identifying suitable regions where significant absorption features from the K-type donor are expected, telluric absorption is reduced and adequate S/N was available. This led us to focus on two key spectral regions; i) 14300-17800\AA~(hereafter referred to as the H-band window) where a number of atomic absorption lines are to be found in a region of reduced telluric absorption and ii) 22200-24800\AA~(K-band window) where the strong CO-bands are located that were previously used to determine the RV of the donor star in \target. Within each region, a number of masks were compared in order to ensure that the exact choice of pixels to mask had no significant impact on the derived radial velocities. In each band, we cross-correlated all 24 target exposures with the 5 spectral template stars. We found that the pipeline did not produce robust uncertainties that would allow us to propagate these uncertainties into a correct formal radial velocity error. This 
is not surprising given the complex nature 
of extracting near-infrared spectroscopy from an echellette-type instrument. Instead, we used our RV fit residuals to determine a robust and realistic RV uncertainty by calculating the covariance matrix derived from weighted least-squares fits. By comparing different spectral regions, pixel masks and independent spectral templates, we also have a good handle on possible systematic effects. 

Initial RV fits using the established spectroscopic orbital period of P=33.5d confirmed the detection of moving donor star features consistent with previous work. In order to establish a more robust orbital period, we calculated Lomb-Scargle normalised periodograms using our 24 velocities covering a baseline of over a year. A strong peak at the orbital frequency was invariably present near $P=33.85$d (see Figure \ref{scargleplot}) with minor variations in the peak position between different templates at the level of $\pm0.01$d.
This is larger than the 33.5d period adopted by G01, though within their quoted uncertainty. 
Hurley et al. (2013) recently revisited the VLT data analysed by G01 as well as some additonal data and also favour a period close to 33.8d.
Our ability to improve upon the previously reported period is corroborated by the fact that using P=33.5d results in significantly poorer RV fits. 
Our RV analysis also rules out the possibility that the 30.8 day
photometric period reported by Neil, Bailyn \& Cobb (2007) is the orbital period. Phase folding our RVs on either this period, or our closest alias, leads to very poor fits.
To determine the binary phase, we used our RV data and allowed the phase zero point (donor star conjunction) as well as the period to vary to assess optimal values and uncertainties on these parameters. This delivered the following ephemeris that we will use throughout this paper to calculate orbital phases:

\[ T_O (HJD(UTC)) = (2455458.68 \pm 0.06) + (33.85 \pm 0.16) E \]

We found that formal fit parameter uncertainties were larger than the variations amongst the 5 sets of RVs using different templates. The particular choice of the latter therefore has no significant impact on the ephemeris.

\begin{table}
\begin{center}
\caption{Radial Velocity Solutions\label{rvtable}}
\begin{tabular}{llcc}
\tableline\tableline
Star & Spectral &  K (km/s) & \\
 ID    &   type       & H-band  & K-band\\     
\tableline
HD~179130 & K2III  & 125.8$\pm$1.5 & 126.8$\pm$1.4\\
HD~174336 & M0III  & 124.7$\pm$1.3 & 126.7$\pm$1.1\\
HD~180732 & K5III & 125.3$\pm$1.1 & 127.0$\pm$1.2\\
HD~179691 & K1III & 125.5$\pm$1.3 & 126.9$\pm$1.3\\
HD~176354 & K0III & 125.6$\pm$1.2 & 127.6$\pm$1.4\\
\tableline
mean & & 125.4$\pm$0.7 & 127.0$\pm$0.6\\
%Giant mean & & 31.7$\pm$0.7 & 32.9$\pm$0.8\\
\tableline
\end{tabular}
%% Any table notes must follow the \end{tabular} command.
%\tablenotetext{a}{The two columns reflect our two choices for the linear limb darkening coefficient}
\end{center}
\end{table}

\begin{figure}
\centerline{\includegraphics[width=9cm]{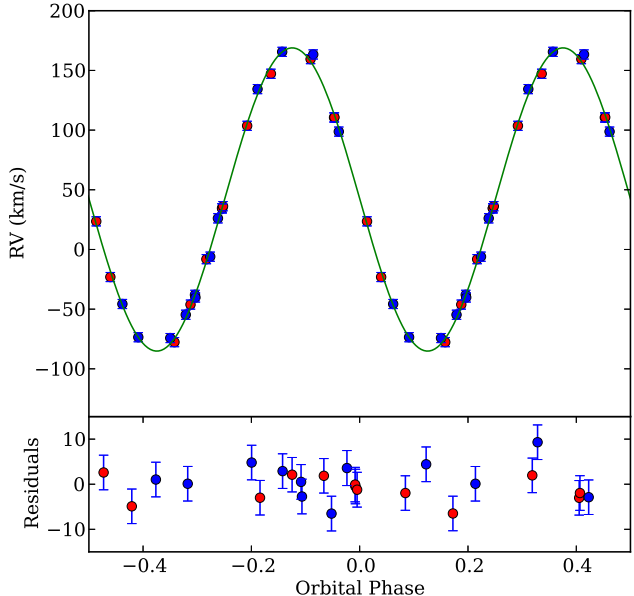}}
\caption{Example radial velocity curve obtained by cross-correlating our 24 \target~spectra with a K5III template in the K-band. Red points denote 2011 data, blue points 2010 data. Data are displayed over two cycles for plotting purposes and error bars are only just larger than the plot symbols.
\label{rvplot}}
\end{figure}

In Figure \ref{rvplot} we show an example RV curve plus fit relative to a K5 template. We see that our 24 epochs achieve good phase coverage across the binary orbit and our fit residuals show a RMS of 3-4 km/s, roughly half a detector pixel. We find that a sinusoidal orbit achieves good fits, suggesting that distorion of the velocities due to heating or irradiation is minimal. In our RV fits, we optimize the mean velocity ($\gamma$), the semi-amplitude ($K$) and the phase zero offset ($\phi_0$) as our fit parameters using a weighted least-squares minimizer, with parameter uncertainties calculated from the covariance matrix. Table \ref{rvtable} lists the resultant fit parameters for our five templates and our two key spectral regions. We see that our semi-amplitude is robustly determined with typical fit uncertainties of 1-1.5 km/s, and that there is excellent agreement between different templates.
We find a marginal difference between the semi-amplitudes derived from the H and K-band, with the H-band delivering slightly lower values. This 
difference is formally not 
significant, though we do note that in the K-band we rely mainly on the CO band-head features, while in the H-band the spectrum consists of a large number of atomic lines (see below). To be conservative, we will include this effect as part of our adopted error on K.
Although we are confident of our orbital period measurement, we verified that when allowing both period and phase zero as additional free parameters, the resultant semi-amplitudes only differ by 0.3 km/s compared to those derived using the above ephemeris. 
This allows us to confidently constrain the projected donor star orbital velocity to $K_2 = 126 \pm 1$km/s, representing a significant improvement over previous work (G01; $K=140\pm15$km/s). This improvement is due to a combination of significantly higher spectral resolution than employed previously, wide wavelength coverage allowing us to include many lines, better S/N per epoch and more extensive phase coverage. The resolution in particular is key to allow for a good sky and telluric correction as well as resolving the narrow donor star features.

Our mean velocity offsets are relative to the template used for the cross-correlation with uncertainties below 1 km/s. For one of our template stars a published radial velocity is available (HD 179130 has a heliocentric radial velocity of $-38 \pm 4.4$ km/s; Wilson 1953, Gontcharov 2006). This allows us to determine the systemic velocity to \target~to be $+11 \pm 4.5$ km/s in the heliocentric frame, with the error dominated by the published RV measurement of our template. This is consistent with G01, but significantly smaller than Hurley et al. (2013), who advocate $\gamma=34.2 \pm 2.5$ km/s. Such a high value is ruled out by our data which have been carefully aligned using the telluric features as described above and is based on a comparison with a radial velocity standard star observed and reduced in an identical manner.

\begin{figure*}
\centerline{\includegraphics[width=16cm]{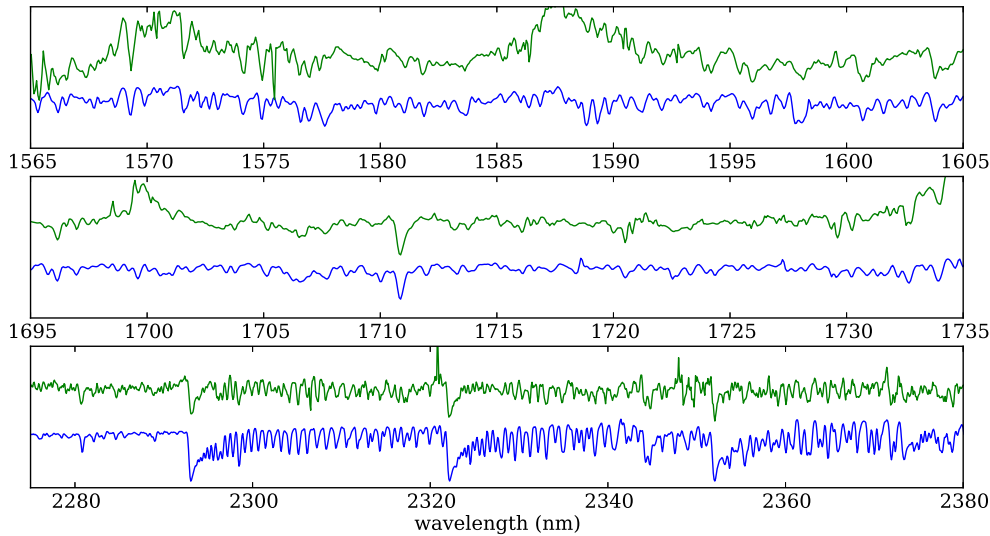}}
\caption{Comparison between the average spectrum of \target~ in the restframe of its mass donor (upper spectrum) and a K5III field giant. Both spectra are continuum normalised and the K5III spectrum has been scaled to reflect the approximate 25\% contribution of the donor star to the total light in \target. A large number of common features can be identified, mainly atomic lines in the H-band and the strong CO band-heads in the K-band. Broad emission features can also be seen in \target, though may be affected by telluric star absorption (see Section \ref{redu}).
\label{specplot}}
\end{figure*}

Having determined the radial velocities for all epochs, we subtracted the radial velocity from each spectrum so as to align all spectra in the rest-frame of the donor star. All spectra were then averaged in order to construct a high S/N spectrum highlighting any donor star features. This confirmed the presence of many absorption features in common between \target~and our templates, as expected given the excellent results from the cross-correlation analysis. In Figure \ref{specplot}, we plot a few representative spectral regions illustrating this match. While the CO band-heads offer the strongest features in the K-band, we also see a large number of atomic absorption lines. This again supports the precision of our RV results as such a large number of features ensures robust cross-correlation functions.

\begin{figure}
\centerline{\includegraphics[width=9cm]{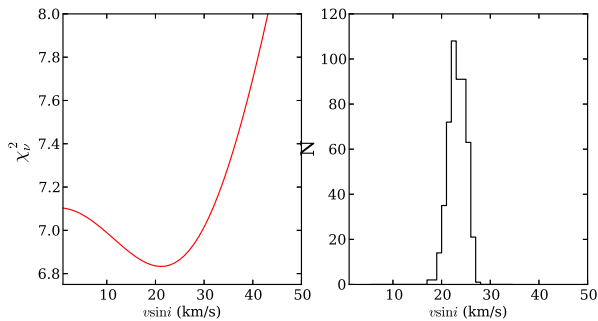}}
\caption{Determination of the rotational broadening of the mass donor star. Top: A scaled and broadened spectrum of the K5III teplate HD~180732 is subtracted from the target spectrum and the $\chi^2$ statistic is used to determine the optimal level of broadening. Bottom: Uncertainties on $v\sin{i}$ are derived from boostrapping the observed spectrum and repeating the optimal subtraction 500 times. The $v\sin{i}$ distribution then allows us to determine confidence limits on our $v\sin{i}$. \label{vsiniplot}}
\end{figure}

Armed with a high S/N spectrum of \target~in its donor rest-frame, we used this to determine the rotational broadening of the donor ($v\sin{i}$) by subtracting a scaled and broadened version of the templates from this average spectrum. The scaling factor reflects the fact that not all of the light originates from the donor, while the amount of applied rotational broadening was varied in order to assess which amount of broadening achieves the best subtraction. Here we used $\chi^2$ as our goodness of fit parameter, by comparing the residuals between this subtracted spectrum and a smoothed version of the original spectrum (Figure \ref{vsiniplot}a). Finally, in order to determine a robust uncertainty on this optimal $v\sin{i}$, we repeated this process using 500 bootstrap copies of the average spectrum. This delivered 500 $v\sin{i}$ values with
a distribution well described by a Gaussian, allowing us to use the mean and rms of this distribution to determine $v\sin{i}$ and its error (Figure \ref{vsiniplot}b). 

Each of the 5 spectral templates in the two wavelength regions provide us with an independent set of measurements. We find optimal values ranging from 17-25 km/s with uncertainties of 2-3 km/s for a given template and spectral region. The donor star appears, on average, to make a 20-25\% contribution to the K-band flux. We note that this small amount of broadening spans 2-3 pixels, but is just below our formal spectral resolution (3.5 pixels). We must therefore be cautious about this measurement despite the consistency among the 5 templates. We have used template stars observed with the same instrument to ensure that the instrumental contribution to the line profile shape is the same. We can also safely ignore the intrinsic $v\sin{i}$ of the templates as for K-giants this is only of the order of 4 km/s (Massarotti et al. 2008).
We note that our values are very close to those of Harlaftis \& Greiner (2004), obtained from significantly lower resolution ISAAC data.  

Our data confirm a low value for the rotational broadening, which in turn implies a low mass ratio for the binary (see next Section), but we cannot rule out that our determination may still be slightly biased as it is just below our spectral resolution (see also Steeghs \& Jonker 2007).  We adopt $v\sin{i}=21\pm4$ km/s as our best estimate, ensuring that our error is large enough to encompass the effects described above. Even higher spectral resolution data would be valuable to reduce this uncertainty.

\section{The mass of the black hole}
\label{mass}

Having determined an accurate value for the donor star's semi-amplitude ($K_2=126\pm1$ km/s) and its rotational broadening ($v\sin{i}=21\pm4$ km/s), the binary mass ratio can be constrained, as the ratio between $K_2$ and $v\sin{i}$ is a function of the mass ratio only (Wade \& Horne 1988). This assumes that the mass donor star is Roche lobe filling and in corotation with the binary. This results in $q=M_{donor}/M_{BH}=0.042\pm0.024$, with the error dominated by the relatively large error on the small $v\sin{i}$. For the expected inclination range discussed belwo, this is equivalent to a donor star radius of 
$R_{donor}=15.6 \pm 2.6 R_{\odot}$, slightly smaller than previous estimates due to our marginally smaller $v\sin{i}$ and consistent with the typical size of a K giant. As for $q$, the uncertainty here is dominated by $v\sin{i}$.

A crucial final step towards a refined determination of the black hole mass requires constraints on the binary inclination. The commonly adopted inclination of $i=66\pm2^{o}$ (Fender et al. 1999) is derived from long-term radio monitoring of relativistic ejecta from \target. However, this value is dependent on the assumed distance, with the inclination dropping if the source is closer than 11 kpc. 
% distance
Distance estimates for \target~cover 6-12 kpc, though $d=11$ kpc is most commonly adopted, as this distance is consistent with a number of independent determinations (see the discussion in Zdziarski et al. 2005). A distance much larger than 12kpc is ruled out as this would imply jet velocities exceeding the speed of light, unless the ejecta are significantly asymmetric (Fender et al. 1999). G01 estimated $d=12.1\pm0.8 $kpc, assuming the systemic velocity of \target~tracks galactic rotation which represents a systematic effect not included in the error estimate. 
If we translate our slightly larger, but more precise, value for $\gamma=+11$km/s into the LSR frame and employ the Reid et al. (2009) Galactic parameters, we find $d=10.4 \pm1.3$ kpc. Here the error allows for a 20 km/s peculiar velocity component, but this may still be too small if the \target~progenitor binary system received a substantial kick when the black hole was formed. We note our formal error on $\gamma$ is already small enough not to dominate the above distance estimate.
% Mark's result
We are also conducting a long-term VLBI campaign in order to determine the distance to \target~via the trigonometric parallax method. Our data to date indicate a parallax of $\pi=0.084 \pm 0.033$ mas, again favouring $d=11-12$ kpc, although we do not yet have the statistics to constrain this firmly (Reid et al., in preparation).
We thus adopt $d=11\pm1$ kpc as our preferred distance range, though we will also explore a more conservative scenario where the distance is allowed to span the full 6-12 kpc range. 
% jet/disk alignment
We acknowledge that using the jet inclination as the binary inclination assumes the inner disc to be co-planar with the binary orbit. This is expected to be a safe assumption for microquasars given their age and the magnitude of the alignment torque (e.g. Steiner \& McClintock 2012).

\begin{figure}
\centerline{\includegraphics[width=10cm]{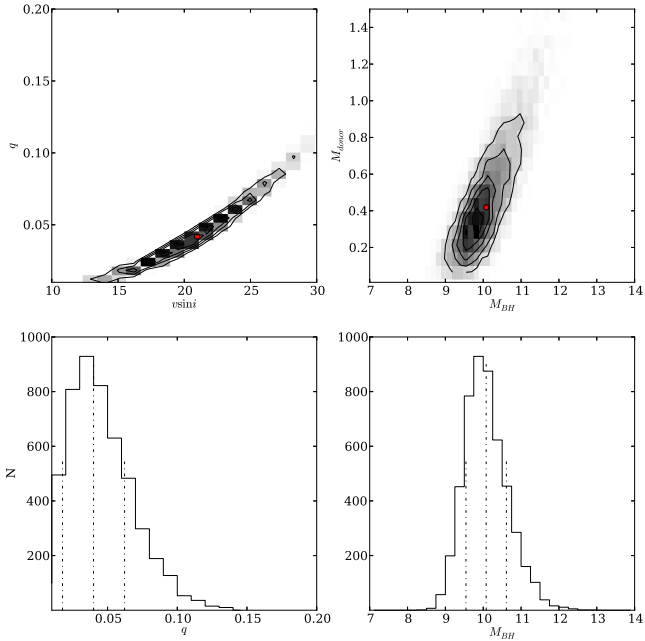}}
\caption{Monte-Carlo calculation of the implied black hole mass given the uncertainties on the four input parameters ($K_2,v\sin{i},P_{orb}$ and $i$). Each pair of $K_2,v\sin{i}$ values delivers a mass ratio (left panels) with the top panel showing that $v\sin{i}$ is the dominant error in the mass ratio determination. Together with the error on $P_{orb}$ and inclination we then calculate the mass distribution for the black hole and its donor star (right panels). In the single parameter histograms on the bottom row, vertical dashed lines denote the mean value and the $\pm1\sigma$ confidence intervals.
\label{massplot}}
\end{figure}

In order to determine the implied black hole mass with a realistic uncertainty, we performed a Monte-Carlo simulation whereby we varied all observed parameters within their expected probability distributions and calculated the black hole mass for 5000 combinations of all four input parameters. Specifically, we use Gauss-normal distributions for $K_2$, $v\sin{i}$ and the binary period using the 1$\sigma$ errors derived from our data. For the inclination, we start by picking a random distance from our adopted distance range and use the Fender et al. (1999) mapping to translate this into a binary inclination. Here we use a random uniform distribution rather than a normal distribution given that we merely assume the distance to lie within a certain range. With all parameters then picked for each Monte-Carlo trial, the underlying calculation simply uses Kepler's law where each $K_2$ and $v\sin{i}$ pair is combined to deliver the mass ratio for that trial as explained above (see Figure \ref{massplot}).

For $d=10-12$ kpc we find $M_{BH}=10.1\pm0.6 M_{\odot}$ and for the donor $M_{2}=0.42\pm0.27 M_{\odot}$. We again stress that the contributions from all individual parameter estimates are included in these mass errors. If we only apply loose constraints on the distance ($d=6-12$ kpc), this results in $M_{BH}=11.7\pm1.8 M_{\odot}$. 
Even in this latter case, the result is a significant improvement
over published estimates, despite these not including the full distance uncertainty.

\section{Conclusions}
\label{conclude}

We have presented a new dynamical study targeting the mass of the black hole in the microquasar \target. We obtained high quality near-infrared spectroscopy aimed at the detection of photospheric absorption features from the mass donor star across a wide wavelength range. The \instrument spectrograph on the VLT allowed us to 
resolve many such features across the H and K-bands at superior resolution and S/N. We detect and resolve, for the first time,  a large number of atomic lines in addition to the CO bandheads that were previsouly used for the radial velocity analysis. 
Our 24 epochs of spectroscopy lead to a significantly improved determination of the orbital parameters. We constrain the binary period to be $P=33.85 \pm 0.16$d, in agreement with the studies of Greiner et al. (2001a) and Hurley et al. (2013), but rule out the period proposed by Neil, Bailyn and Cobb (2007). 

Cross-correlation analysis with a sample of K-giant template stars provided us with accurate radial velocities, leading to a precise determination of the projected orbital velocity of the mass donor star ($K_2=126\pm1$ km/s). This velocity was robust against choice of template star and wavelength range. Since our spectra cover a large number of atomic lines, we do not rely on molecular bandheads as was done previously, although we find that the two types of features give consistent results, suggesting that at our spectral resolution the use of CO band-heads poses no problems.  
The systemic velocity of the binary system relative to the heliocentre was determined to be $\gamma=+11\pm4.5$km/s by comparison with a radial velocity standard. If we assume that this motion tracks the galactic rotation curve, a distance of $10.4\pm1.3$ kpc is implied.
We combined our individual spectra to construct a high S/N spectrum in the restframe of the mass donor star. This allowed us to determine the rotational broadening of the donor $v\sin{i}=21\pm4$ km/s, equivalent to a radius of $R_{donor}=15.6 \pm 2.6 R_{\odot}$, as well as the binary mass ratio $q=0.042 \pm 0.024$. 

Combining all parameters allows us to constrain the black hole mass tightly, with the key remaining uncertaintly being the assumed binary inclination. Using the commonly adopted distance and inclination, we find $M_{BH}=10.1\pm0.6 M_{\odot}$. A more conservative calculation based on only loose distance constraints leads to $M_{BH}=11.7\pm1.8 M_{\odot}$. Our new mass determination brings the black hole mass of \target~closer in line with core collapse simulations and the stellar BH mass distribution (Ozel et al. 2010; Farr et al. 2001). 

The implied mass for the donor ($M_{2}=0.42\pm0.27 M_{\odot}$) is quite small and lower than the typical mass of a K/M field giant. Although the spectral characteristics of the donor star in \target~appear to be consistent with a K/M giant, it is not uncommon to find under-massive secondaries in accreting binaries due to the evolved nature of such systems (e.g. Casares et al. 2010).

Our analysis shows that near-infrared spectroscopy can offer dynamical studies of obscured systems in the galactic plane and bulge with precisions on par with those of typical optical studies.


%% Included in this acknowledgments section are examples of the
%% AASTeX hypertext markup commands. Use \url without the optional [HREF]
%% argument when you want to print the url directly in the text. Otherwise,
%% use either \url or \anchor, with the HREF as the first argument and the
%% text to be printed in the second.

\acknowledgments

DS acknowledges support from STFC through an Advanced Fellowship (PP/D005914/1)
as well as grant ST/I001719/1. JEM acknowledges the suppport of NASA grant NNX11AD08G.
Based on observations made with ESO Telescopes at the La Silla Paranal Observatory under programme ID 085.D-0497. We thank Tom Marsh for the use of his Molly spectral analysis package. 

\vspace{0.5cm}


%% in parentheses, after the keyword, but they will not be verified.

{\it Facilities:} \facility{ESO (VLT)}

\end{document}